# Quantification of Morphological Features in Non-Contrast Ultrasound Microvasculature Imaging

Siavash Ghavami*[1], Mahdi Bayat**, Mostafa Fatemi**, Azra Alizad*[#2]

* Department of Radiology, Mayo Clinic College of Medicine and Science, Rochester, MN, 55905, USA
**Department of Physiology and Biomedical Engineering, Mayo Clinic College of Medicine and Science, Rochester, MN, 550905, USA

Emails:{[1]roudsari.seyed, [2]alizad.azra}@mayo.edu

#corresponding author

**Abstract-** Morphological features of small vessels provide invaluable information regarding underlying tissue, especially in cancerous tumors. This paper introduces methods for obtaining quantitative morphological features from microvasculature images obtained by non-contrast ultrasound imaging. Those images suffer from the artifact that limit quantitative analysis of the vessel morphological features. In this paper we introduce processing steps to increase accuracy of the morphological assessment for quantitative vessel analysis in presence of these artifact. Specifically, artificats are reduced by additional filtering and vessel segments obtained by skeletonization of the regularized microvasculature images are further analyzed to satisfy additional constraints, such as diameter, and length of the vessel segments. Measurement of some morphological metrics, such as tortuosity, depends on preserving large vessel trunks that may be broken down into multiple branches. We propose two methods to address this problem. In the first method, small vessel segments are suppressed in the vessel filtering process via adjusting the size scale of the regularization. Hence, tortuosity of the large trunks can be more accurately estimated by preserving longer vessel segments. In the second approach, small connected vessel segments are removed by a combination of morphological erosion and dilation operations on the segmented vasculature images. These methods are tested on representative *in vivo* images of breast lesion microvasculature, and the outcomes are discussed. This paper provides a tool for quantification of microvasculature image from non-contrast ultrasound imaging may result in potential biomarkers for diagnosis of some diseases.

# I. INTRODUCTION

Microvasculature architecture is known to be associated with tissue state and pathology. Different conditions and diseases can alter vasculature at different size scales. Several studies have shown that malignant tumor growth coincides with changes in the vascularity of normal tissue [1-3]. Malignant tumors are known to present different mechanical features leading to the growth of more permeable and tortuous vessels [4, 5]. Vessel tortuosity has been found to reveal information about some diseases [6, 7]. Moreover, it has been shown that microvascular parameters such as vessel size and branching correlate very well with tumor aggressiveness and angiogenesis [8].

Several preclinical and clinical studies are available to derive quantitative information from microvasculature images obtained by contrast agent ultrasound for diagnostic purposes [9-13]. Additional examples include perfusion imaging [14, 15] and molecular imaging [16, 17]. Conventionally, these techniques endeavor to screen either the measure of blood flow inside a tissue volume by testing the increase in ultrasound signal from the blood pool contrast agents, or the presence of molecular markers of an ailment through imaging of the targeted contrast agent held in the blood flow. While a few researches have recently shown the capacity to quantify the architecture of the blood vessels in thyroid nodules and breast lesions, the use of contrast agents remains a barrier for extensive investigations [18-20]. On the other hand, analyzing vascular networks using ultrasound imaging devoid of contrast agent is a new framework made possible only recently, thanks to new clutter removal processing methods [21-23]. For example, Cohen *et al.* [24] demonstrated that vascular structures provide useful information for neuro-navigation in brain imaging. This framework exploits the coherence of the tissue data provided by fast plane wave imaging of a large field of views to enable detailed imaging of the micro-vasculature structure by integrating longer data ensembles.

Blood vessel segmentation and analysis techniques have been studied exhaustively in other imaging modalities, such as optical imaging of the retina [25]. Retinal vessel segmentation algorithms are a principal component of automatic retinal infection screening frameworks. Different vessel analysis methods used in retinal images acquired by a fundus camera and have been summarized in detail in the literature [25]. Application of the brain vessel segmentation has also been described in magnetic resonance imaging (MRI) [26]. A vessel analysis tool was reported for morphometric measurement and representation of the vessels in computed tomography (CT) and MRI data sets [27]. Methods of blood vessel segmentation algorithms have been reviewed widely in the literature [28]. An isotropic minimal path-based framework has been proposed for segmentation and quantification of the vascular networks [29].

In this paper, we focus on challenges of vessel quantification for 2-dimensional (D) label-free ultrasound Doppler imaging and propose solutions to overcome such challenges. We evaluate performance of proposed solutions on the quantification of in vivo breast data. As noted, vessel quantification has been widely used in a broad range of imaging modalities; however, adaptation for the analysis of the microvasculature images obtained by non-contrast ultrasound requires careful treatment. This imaging modality, while enabling a versatile mechanism for acquiring small vessel images, introduces some challenges. The main problem stems from the 2-D interpretation of 3-D vascular structures. The work in retinal vessel analysis, while performed in 2-D, only considers surface vascularity for which a 2-D model is well-justified. When considering vessels distributed in a volume, 2-D cross-sectional ultrasound imaging may provide erroneous branching and vessel crossings that can lead to incorrect interpretation of vessel segments. While quantitative evaluation of parameters such as vessel density and diameter are not significantly affected by this phenomenon, measures of tortuosity, branching points, and number of vessel segments may become inaccurate. Another difficulty arises from imaging vessels in the cross-sectional orientation. Vessels may appear as small segments with incorrect information regarding vascular tree segments. The main contribution of this paper is to address these issues using either vessel filtering or morphological operations such that the most dominant vascular features can be obtained from 2-D non-contrast ultrasound imaging.

The remainder of this paper is organized such that Section II contains methods which include a background on image formation, Hessian-based filtering, vessel segmentation, morphological filtering, vessel quantification, and patient studies. Results are presented and discussed in section III. Finally, in section IV offers our main conclusions and future directions.

## II. MATERIALS AND METHODS

Before quantitative information of vessels can be extracted from ultrasound microvasculature images, one must perform multiple preprocessing steps. The first step is image formation, which reconstructs the microvasculature image from a sequence of plane wave ultrasound images [22]. Second, vessel filtering is used to enhance the structure of vessels and provide adequate background separation for segmentation. Morphological filtering, vessel segmentation, and skeletonization occurs last. The main contribution of this paper is in the last two modules, i.e. vessel segmentation and filtering and vessel quantification. First, though, we provide background on image formation and Hessian-based filtering.

### A. Background on image formation

Figure 1 demonstrates the sequence of processing steps of the microvasculature image formation algorithm with an example of output image in each processing step. Processing begins with storage of ultrasound plane wave data in in-phase–quadrature (IQ) format. This data can be characterized by the complex-valued variable $s(x,z,t)$, where $x$ and $z$ denote the lateral and axial dimensions, respectively, and $t$ denotes the ultrasound imaging slow time. This signal can be described as the sum of three components as follows

$$s(x,z,t) = c(x,z,t) + b(x,z,t) + n(x,z,t), \tag{1}$$

where $c(x,z,t)$, $b(x,z,t)$ and $n(x,z,t)$ represent the clutter signal, the blood signal, and the additive thermal noise, respectively. Spatial and temporal characteristics of these three components differ, and $n(x,z,t)$ is considered as zero mean Gaussian white noise. Signal $s(x,z,t)$ corresponds to tensor $\mathbf{S} \in \mathbb{R}^{n_x \times n_z \times n_t}$, where $n_x$ and $n_z$ are the number of spatial samples along the x-direction and z-direction, respectively, and $n_t$ is the number of samples over time. The data tensor $\mathbf{S}$ is reshaped to form a Casorati matrix by transforming tensor $\mathbf{S}$ into a 2-D spatiotemporal matrix $\mathbf{S}_C \in \mathbb{R}^{(n_x \times n_z) \times n_t}$. This transformation has been also proposed in other imaging modalities like MRI and CT [30-33].

Using singular value decomposition (SVD) of $\mathbf{S}_C$ we have [34]

$$\mathbf{S}_C = \mathbf{U}\boldsymbol{\Delta}\mathbf{V}^{\dagger} \qquad (2)$$

where $\boldsymbol{\Delta} \in \mathbb{R}^{(n_x \times n_z) \times n_t}$ is a non-square diagonal matrix, $\mathbf{U} \in \mathbb{R}^{(n_x \times n_z) \times (n_x \times n_z)}$ and $\mathbf{V} \in \mathbb{R}^{n_t \times n_t}$ are orthonormal matrices, and $\dagger$ indicates conjugate transpose. Columns of $\mathbf{U}$ and $\mathbf{V}$ matrices correspond to the spatial and temporal singular vectors of $\mathbf{S}_C$. Based on the definition of SVD, matrix $\mathbf{S}_C$ can be decomposed into sum of rank one matrix $\mathbf{A}_i = U_i \otimes V_i$ as follows:

$$\mathbf{S}_C = \sum_i \lambda_i \mathbf{A}_i = \sum_i \lambda_i U_i \otimes V_i \qquad (3)$$

where $U_i$ and $V_i$ are $i^{\text{th}}$ columns of $\mathbf{U}$ and $\mathbf{V}$, respectively, $\lambda_i$ is $i^{th}$ ordered singular values of $\mathbf{S}_C$ and $\otimes$ denotes outer product operation. Each column of $V_i$ is a temporal signal with length $n_t$. Each column $U_i$ is spatial signal with dimensionality of $n_x \times n_z$. In fact, each vector of $U_i$ describes a 2-D spatial image $I_i$ which is modulated by a temporal signal $V_i$. Hence, we have

$$s(x,z,t) = \sum_{i=1}^{rank(S_c)} \lambda_i I_i(x,z) V_i(t). \qquad (4)$$

In low rank clutter filtering framework, tissue component is considered to comprise the first few dominant singular values and vectors, while blood signal is formed by the subsequent singular values when sorted in a descending order. Based on these assumptions on tissue and blood signals, clutter removal is performed using a threshold $n$ on the number of singular vectors removed from $s(x,z,t)$. Therefore, blood signal is derived as follows

$$s_{blood}(x,z,t) = s(x,z,t) - \sum_{i=1}^{n} \lambda_i I_i(x,z) V_i(t). \qquad (5)$$

In this paper threshold $n$ is selected based on setting a threshold on the slope of the second order derivative of eigenvalues decay, as described by Bayat, et al. in [22]. The filtered signal $s_{blood}(x,z,t)$ is used to produce the power Doppler image as

$$I(x,z) = \sum_{k=1}^{K} |s_{blood}(x,z,kT)|^2 \qquad (6)$$

where $T$ is the sampling time between two successive ultrafast ultrasound frames. To further enhance clutter removal performance, an additional step can be introduced before forming the intensity image in (6). This additional step enforces unilateral Doppler shift which is expected to occur from the unidirectional flow in vessels. Hence, the final image can be formed as

$$I(x,z) = |I_p - I_n| \qquad (7)$$

where $I_p$ is the energy at the positive frequency side of the spectrum and is defined as

$$I_p = \int_0^{\infty} |S_{blood}(x,z,f)|^2 \, df \qquad (8)$$

where $S_{blood}(x,z,f)$ is Fourier transform of $s_{blood}(x,z,t)$ and $I_n$ is energy at the negative frequency side of the spectrum and is defined as

$$I_n = \int_{-\infty}^{0} |S_{blood}(x,z,f)|^2 \, df. \qquad (9)$$

Finally, a top hat filter (THF) is applied on $I(x,z)$ to remove the background noise. A THF is comprised of a background estimation, followed by a background subtraction, operation [35]. Details about application of this filter for background removal of non-contrast ultrasound microvasculature images has been previously described [22]. The output image of THF is denoted by $I_T(x,z)$.

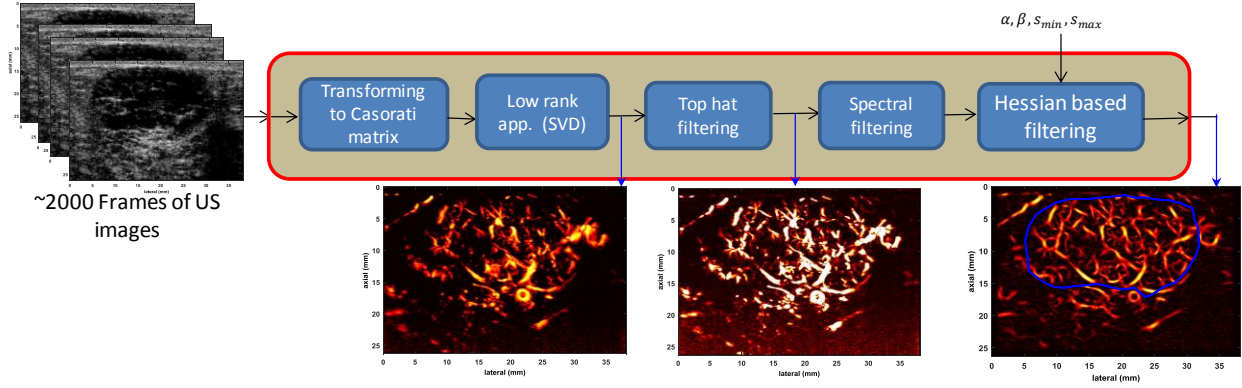

Figure 1. Block diagram of microvasculature image formation module.

*B. Hessian-based filtering*

To enhance the visibility of the microvasculature image in the presence of strong background signals, morphological filtering based on a THF was used. Due to background noise, though, random patterns will also be present at the output of a THF. Hence, vessel enhancement filters are used to penalize background noise and further enhance vessel structure. Enhancement filters based on the analysis of eigenvalues of the Hessian matrix applied on a $2$-D image selectively amplify a specific local intensity profile or structure in an image. Hessian-based filters [36] distinguish between different local structures by analyzing the second order intensity derivatives at each point in the image. To enhance the local structures of various sizes, the analysis is typically performed on a Gaussian scale space of the image as previously described [22].

The image output of THF, $I_T(x,z)$ denotes the intensity of a 2-D image at coordinates $(x,z)$. The Hessian of $I_T(x,z)$ at scale $s$ is then represented by a $2\times 2$ matrix:

$$H(x,z,s) = \begin{bmatrix} s^2 I_T(x,z) * \dfrac{\partial^2}{\partial x^2} G(x,z,s) & s^2 I_T(x,z) * \dfrac{\partial^2}{\partial x \partial z} G(x,z,s) \\ s^2 I_T(x,z) * \dfrac{\partial^2}{\partial z \partial x} G(x,z,s) & s^2 I_T(x,z) * \dfrac{\partial^2}{\partial z^2} G(x,z,s) \end{bmatrix} \quad (10)$$

where $s$ is size scale of filtering, $G(x,z,s) = (2\pi s^2)^{-1} \exp(-(x^2+z^2)/2s^2)$ is a 2-D Gaussian function and $*$ denotes convolution.

Selective enhancement of the local image structures is independent of the orientation. This enhancement is based on shape and brightness of the structures and can be done by analyzing the signs and magnitudes of the Hessian eigenvalues i.e. $\lambda_i \ \ i \in \{1,2\}$. At each point $(x,z)$, the eigenvalues are obtained via eigenvalue decomposition of $H(x,z,s)$.

Ideally, the Hessian-based enhancement maps eigenvalues of $H(x,z,s)$ to values 0 or 1. Therefore, this filter can be considered as a response of an indicator function of a certain set of eigenvalue relations. To limit unwanted fluctuations due to intensity variations of ultrasound Doppler image or noise, the indicator functions are approximated by smooth enhancement functions $\upsilon:[\text{eig} H(x,z,s)] \to R^+$ which have a non-negative response. Since vessel orientation is not of interest in the analysis presented in this paper, the eigenvalues, $\lambda_1$ and $\lambda_2$ (assuming $|\lambda_1| > |\lambda_2|$), are only considered for additional vessel analysis. Noise-like variations do not change eigenvalues significantly. Consequently, the energy of the eigenvalues can be used as a measure for assessment of the structured shapes (e.g. tubes and blubs), as oppose to random patterns. To minimize effect of unwanted fluctuation of ultrasound signal, a multiscale filter response $F(x,z)$ is then obtained by maximizing a given enhancement function $\upsilon$, at each point $(x,z)$, over a range of scales $s$ as follows:

$$F(x,z) = \sup_s \{\upsilon[\text{eig} H(x,z,s)] : s_{\min} \leq s \leq s_{\max}\} \tag{11}$$

where $\sup\{\cdot\}$ denotes supremum and the values of $s_{\min}$ and $s_{\max}$ are selected according to the respective minimal and maximal expected size of the structures of interest.

For 2-D images the following vessel likeliness measure has been proposed[36] and used[22] for vessel filtering of the ultrasound microvasculature images:

$$\upsilon_o(s) = \begin{cases} 0 & \text{if } \lambda_2 > 0 \\ \exp\left(-\frac{R_B^2}{2\beta^2}\right)\left(1 - \exp\left(-\frac{s^2}{2\alpha^2}\right)\right) & \text{otherwise,} \end{cases} \tag{12}$$

where, $R_B = \lambda_2/\lambda_1$ is the blobness measure in 2-D image and accounts for the eccentricity of the second order ellipse; $\alpha$ and $\beta$ are filter parameters. For consistency in notation, output image of Hessian filter is denoted by $I_H(x,z)$.

*C. Morphological filtering, vessel segmentation and skeletonization*

Morphological filtering and subsequent vessel segmentation is performed on the output of the Hessian filter. Morphology analysis is a broad set of image processing operations that process images based on shapes (see Figure 2 and Table 1). Morphological operations apply a structuring element to an input image, creating an output image of the same size. In a morphological operation, the value of each pixel in the output image is based on a comparison to the corresponding pixel in the input image with its neighbors. By choosing the size and shape of the neighborhood, we can construct a morphological operation that is sensitive to the specific shapes in the input image.

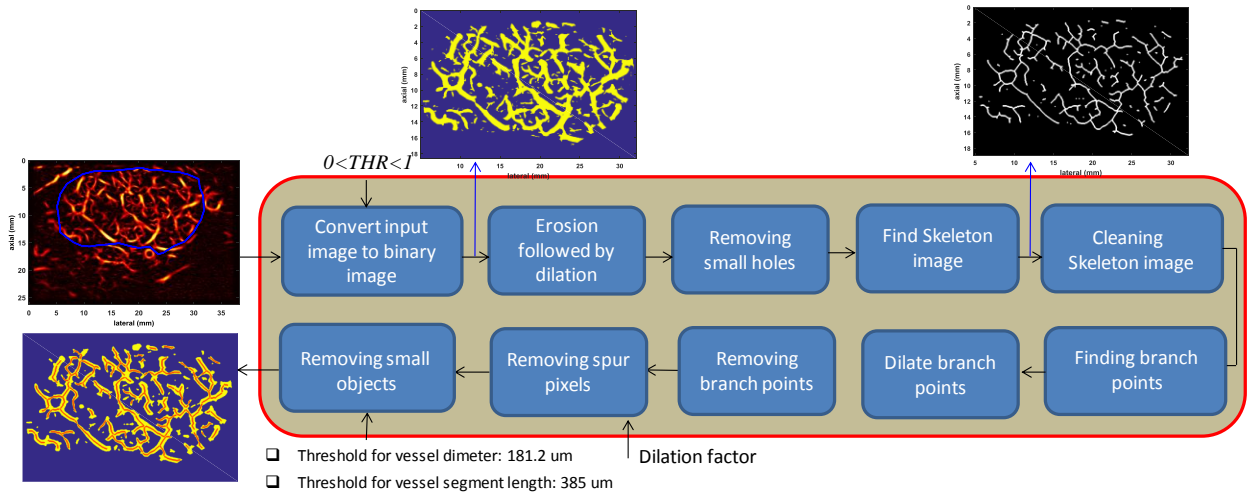

Figure 2. Block diagram of morphological filtering and vessel segmentation module.

Morphological filtering and vessel segmentation include the following steps: converting the microvasculature image (output of Hessian filter) to a binary image, removing small noise-like objects through an erosion and dilation operation, removing small holes, filling small holes with a dilation and erosion operation, finding image skeleton, cleaning skeleton image, removing spur pixels, labeling connected components, finding branch points, dilating branch points, removing branch points, and removing small objects. In the remainder of this section, each step will be briefly described. After these steps, the output image includes the vessel segments. Those segments are analyzed in the vessel quantification module to estimate desired quantitative parameters of the vessels.

Table 1. Pseudo-code of morphological operation

```
Begin
        Converting grayscale image to binary image (C-I)
        Eroding following by dilating (C-II)
        Removing small holes (C-III)
        Finding skeleton image (C-IV)
        Removing isolated pixels
        Removing spur pixels
        Finding branch points
        Dilating branch points
        Removing branch points
        Labeling connected components
        Removing small objects
            • Removing vessels with length less than a threshold
            • Removing vessels with diameter less than a threshold
End
```

*1) Converting grayscale to binary:* The input image to the morphological filtering and vessel segmentation module is the output of Hessian filter and spectral subtraction, i.e. $I(x,z)$, which has values in the [0, 1] range. First, it is converted to a grayscale image. A binary image is then obtained by setting an intensity threshold. The output image $I_B(x,z)$ replaces all pixel values in the input image with luminance $> THR$ with 1 (white), and replaces all other pixel values with 0 (black) where *THR* is a global thresholding value in the range [0, 1].

*2) Erosion followed by dilation:* One of the basic morphological operations is dilation and erosion. Dilation adds pixels to the boundaries of objects in an image, while erosion removes pixels on object boundaries. The number of pixels added or removed from the objects in an image depends on the size and shape of the structuring element used to process the image. In the morphological dilation and erosion operations, the value of any given pixel in the output image is determined by applying a rule to the corresponding pixel and its neighbors in the input image. The rule used to process the pixels defines the operation as dilation or erosion [37]. We use erosion followed by dilation to remove some noise-like small objects in the image after amplitude thresholding.

*3) Removing small holes:* The spectral subtraction in equation 7, while providing additional clutter suppression, may induce erroneous intensity nulling in the image, at isolated points along the vessels with horizontal orientation. This intensity nulling is the effect of spectral nulling due to the ultrasound beam being perpendicular to the blood flow, resulting in symmetric spectrum in the frequency domain. Thus, the two components in spectral subtraction of equation 7 cancel each other. To avoid erroneous splitting of the vessels at these points, a morphological "hole-filling' step is added. This step sets a pixel to 1 if five or more pixels in its 3-by-3 neighborhood are 1s; otherwise, it sets the pixel to 0. After this operation, some small holes may still remain. To remove the remaining small holes in the vessels, we use the operation of dilation followed by erosion

*4) Finding the skeleton image:* The next step in the morphological operations on binary images is removing pixels so that an object without holes shrinks to a line, and an object with holes shrinks to a connected ring halfway between each hole and the outer boundary. Finding the skeleton image is based on a thinning algorithm [38]. Figure 3 (a) shows an example of a binary vessel segment used as the input image of a thinning algorithm. Figure 3 (b-d) show the output image of a thinning algorithm after 5, 10 and 20 iterations of the algorithm. By increasing the number of iterations, the output image of the thinning algorithm converges to a skeleton image. When we select an infinite number of iterations, the iterations are repeated until the image stops changing.

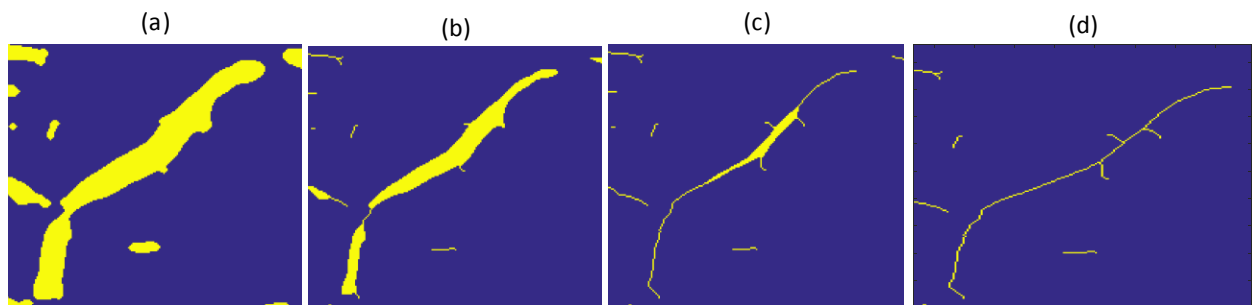

Figure 3: (a) Input vessel image, (b) Output image after 5 iterations, (c) 10 iterations, and (d) 20 iterations of a thinning algorithm

D.  *Vessel quantification*

Vessel segments (overlaid on the binary vessel image in Figure 4) are used for vessel quantification. Figure 4 shows the block diagram of the vessel quantification algorithm. We use three types of vessel analysis: structure analysis, diameter analysis, and tortuosity analysis. Therefore, the quantification parameters considered here include the number of vessel segments, vessel density, the number of branch points, the diameter of vessels, and vessel tortuosity. Two different tortuosity metrics are considered: the distance metric (DM) and the sum of angle metric (SOAM). We use Moore-Neighbor tracing algorithm modified by Jacob's criteria [39] to track vessels. The location vector of vessel $j$ is $\mathbf{P}_j := \left[ \mathbf{p}_{1j}, ...., \mathbf{p}_{N_j j} \right]$, where $\mathbf{p}_{i,j} := \left[ x_{ij}, z_{ij} \right]^T$ is the point $i$ in the vessel $j$, and $N_j$ is the length of vessel $j$.

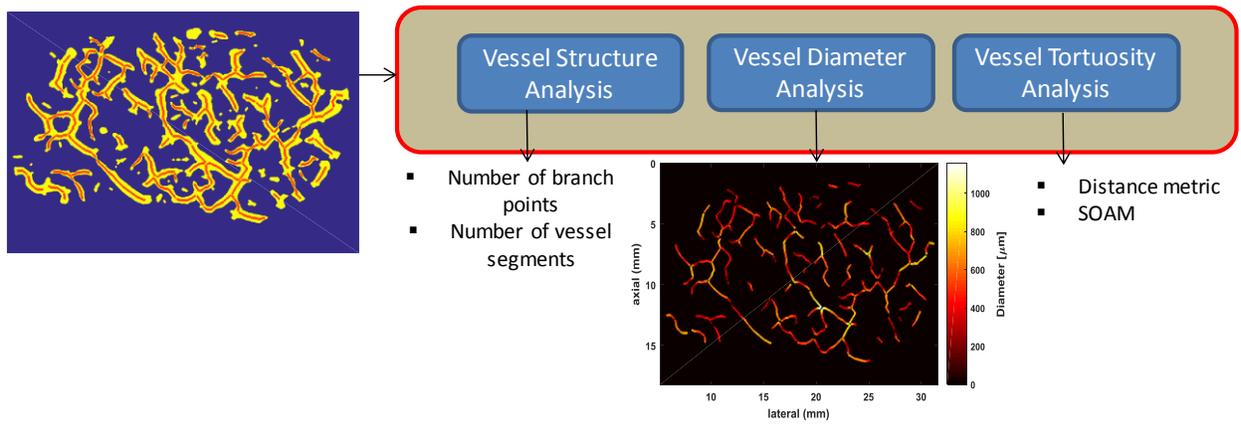

Figure 4. Block diagram of vessel quantification module. SOAM, sum of angle metric.

*1) Distance Metric" (DM):* The DM of vessel $j$ is the most common approach used to evaluate vascular tortuosity in 2-D [25] . It provides a ratio between the actual path length of a meandering curve and the linear distance between endpoints as depicted by

$$DM_j = \frac{\sum_{k=2}^{N} \left| \mathbf{p}_{k,j} - \mathbf{p}_{k-1,j} \right|}{\left| \mathbf{p}_{N,j} - \mathbf{p}_{1,j} \right|}. \tag{13}$$

*2) Sum of Angle Metric (SOAM):* The displacement vectors between points $\mathbf{p}_{k-1,j}$, $\mathbf{p}_{k,j}$ and $\mathbf{p}_{k+1,j}$, $\mathbf{p}_{k,j}$ in vessel $j$ are defined by

$$\mathbf{d}_{k,j} = \mathbf{p}_{k,j} - \mathbf{p}_{k-1,j} \tag{14}$$

$$\mathbf{d}_{k+1,j} = \mathbf{p}_{k+1,j} - \mathbf{p}_{k,j} \tag{15}$$

$$\mathbf{d}_{k+2,j} = \mathbf{p}_{k+2,j} - \mathbf{p}_{k+1,j} \tag{16}$$

where $k \in \{2,...,N_j - 2\}$ and $N_j$ is the length of the vessel in pixels. The in-plane angle at point $\mathbf{p}_{k,j}$, is given by

$$I_{kj} := \cos^{-1}\left(\left(\frac{\mathbf{d}_{k,j}}{|\mathbf{d}_{k,j}|}\right) \cdot \left(\frac{\mathbf{d}_{k+1,j}}{|\mathbf{d}_{k+1,j}|}\right)\right) \tag{17}$$

The torsional angle at point $\mathbf{p}_{k,j}$ is represented by the angle between the plane of the current osculating circle, whose surface normal is the normalized cross product of the vector $\mathbf{d}_{k,j}$ and $\mathbf{d}_{k+1,j}$, and the surface of subsequent osculating plane, whose surface normal is the normalized cross product of the vector $\mathbf{d}_{k+1,j}$ and $\mathbf{d}_{k+2,j}$, which is defined as

$$T_{kj} := \cos^{-1}\left(\left(\frac{\mathbf{d}_{k,j} \times \mathbf{d}_{k+1,j}}{|\mathbf{d}_{k,j} \times \mathbf{d}_{k+1,j}|}\right) \cdot \left(\frac{\mathbf{d}_{k+1,j} \times \mathbf{d}_{k+2,j}}{|\mathbf{d}_{k+1,j} \times \mathbf{d}_{k+2,j}|}\right)\right). \tag{18}$$

Since we are performing 2D imaging, 2 components of internal product are parallel and $T_{kj}$ is derived zero or 180. For all analyses in this paper, all torsional angles are considered 0 when they are 180. Therefore, the total angle $CP_{kj} := \sqrt{I_{kj}^2 + T_{kj}^2}$ at point $\mathbf{p}_{kj}$ and vessel $j$ is given by

$$CP_{kj} = |I_{kj}|. \tag{19}$$

The SOAM calculates the total tortuosity of vessel $j$ and is defined as [6]

$$SOAM_j = \frac{\sum_{k=2}^{N_j-2} CP_{kj}}{\sum_{k=2}^{N} |\mathbf{p}_{kj} - \mathbf{p}_{k-1j}|}. \tag{20}$$

*3) Estimating diameter:* To acquire localized vessel diameter, we first invert the binary image. Therefore, pixels corresponding to vessel are 0 and pixels corresponding to background are 1. Next we obtain the Euclidean distance in the inverted image between zero pixels (corresponding to vessel segments) and the nearest non-zero pixel (corresponding to background) of the image. For all pixels corresponding to vessels (zero pixels), the distance to the nearest non vessel pixel (one pixel) is dedicated to that pixel. The set of points inside of the vessel region and the background region is denoted by $V$ and $B$, respectively. For any point of $(x,z) \in V$, the Euclidian distance between $(x,z)$ and all points $(x_b, z_b) \in B$ is calculated and the minimum distance value is obtained as follows:

$$d(x,z) = \min_{(x_b, z_b)} \sqrt{(x-x_b)^2 + (z-z_b)^2} \\ \text{s.t. } (x_b, z_b) \in B \tag{21}$$

Then the image is skeletonized using the thinning algorithm so that the distances along the centerlines can be calculated. The $i^{th}$ point at center line of vessel $j$ is denoted by $(x_{ij}, z_{ij})$. Vessel diameter is simply obtained by doubling the radius value of $d(x_{ij}, z_{ij})$. Therefore, we can estimate the diameter of each vessel at each point by

$$D(x_{ij}, z_{ij}) = 2d(x_{ij}, z_{ij}). \tag{22}$$

For each vessel segment, the average diameter of the vessel segment over points related to that vessel is obtained as follows:

$$D_j = \frac{1}{N_j} \sum_{i=1}^{N_j} D(x_{ij}, z_{ij}). \tag{23}$$

*4) Quantification of vessel trunks:* In 2-D imaging of 3-D vascular structures, some vessels are only partially visible in the imaging plane. Moreover, it is possible that vessels may appear to cross each other when they do not actually cross in 3-D space. This occurs because of the slice thickness of an ultrasound image. The vessels seem to cross if both are within the slice thickness of B-mode and they are not parallel. Most often, one vessel goes out of the imaging plane, making it look like a small branch. One of the consequences of the branching for vessel quantification is that the main trunk breaks into small vessel segments, which may adversely impact quantification of the morphological features of the trunk. To resolve this problem, we propose two strategies and compare the results: (1) Hessian-based filtering with different minimum size scales, and (2) morphological operations to recover large trunk segments after branching of the small vessel segments. In the first method, minimum size scale of the Hessian-based filtering, i.e. $s_{\min}$, controls the formation of small vessels in the image. In the second method, we create a disk-shaped structuring element with radius $r$ μm. Morphological operations using disk approximations run much faster when the structuring element uses approximations. We perform erosion followed by dilation, using the same structuring element for both operations, (i.e, disk-shaped structuring element). We define the erosion/dilation (ED) factor as follows

$$ED := r \tag{24}$$

to remove small objects and analyze trunks inside the lesion. Using this method, vessel structures with a size less than $r$ are removed from the image. In dilation, only structures larger than $r$ that remain in the image are dilated and converted to their original size. Therefore, we expect only vessel trunks to appear in the final image. In the tortuosity analysis, our goal is to analyze the vessels that are fully located in the imaging plane. Therefore, by removing small vessel segments connected to main vessel trunks, it is possible to analyze the main vessel trunk.

*5) Contribution of small vessel segments in tortuosity analysis:* The microvasculature image is constructed from a sequence of 2-D ultrasound plane wave images in which some vessels are only partially visible in the imaging plane. This, in turn, results in observing small vessel segments in the image. The residual noise, when passed through the Hessian-based filtering, might also result in structures that may be perceived as small vessel segments. Hence, an additional step is required to remove unwanted erroneous or partial vessel segments. This is accomplished by enforcing a minimum vessel segment length as part of the quantification tool. This operation alone can considerably change some morphometric values (e.g., DM tortuosity), as small vessel segments may skew the distribution of such morphometric values with no added information.

*E. Patient demographics and histopathological outcomes*

This study was done to demonstrate the process and potential application of microvasculature quantification in vivo. Under an institutional review board-approved protocol, 4 patients with breast lesions participated in this study; they gave written, informed consent beforehand. To assess the performance of the proposed methods for morphological analysis of the microvasculature images obtained by contrast-agent-free ultrasound, an Alpinion Ecube12-R ultrasound machine (ALPINION Medical Systems, Seoul, Korea) and a linear array transducer L3-12H (ALPINION Medical Systems, Seoul, Korea) were used. For each patient, 3 seconds of high frame rate, 5-angle compounded plane wave imaging data were acquired at 680 frames per second. The lesions from the 4 patients were manually segmented using the B-mode image obtained from the first frame in the imaging sequence. All patients underwent biopsy following the ultrasound examination and pathology results were used as the final diagnosis. Vasculature images were obtained using SVD clutter removal filtering, followed by THF background removal and Hessian-based vessel filtering as described previously in the literature [22]. The THF was employed using a disk structuring element of size 577.5 μm. Vessel filtering was applied using size scales in the range of 115.5 μm to 346.5 μm. Values outside this range are explicitly noted in the results section. The vessel's filter parameters $\alpha$ and $\beta$ were set to 1 and 0.6, respectively. Vessel images were further analyzed to acquire morphological parameters using the proposed method. The minimum length for a vessel segment was considered 385 μm; the minimum diameter for a vessel segment was considered to be 181.2 μm. To convert the gray scale images to binary, a *THR* of 0.15 was used, which is obtained empirically for noise fluctuations removal.

## III. RESULTS

We applied the vessel quantification algorithm on the microvasculature images obtained from vascularized breast lesions. Four different lesions: 2 malignant [invasive/infiltrating ductal carcinoma (IDC) grade III], and 2 benign (fibroadenomas) were studied. For clarity, we refer to them as IDC grade III (1) and (2) and fibroadenoma (3) and (4). We derived quantitative parameters of the microvasculature images for these 4 lesions to study and address challenges of vessel quantification using contrast-agent-free ultrasound imaging.

Figure 5 (a-d) depict the steps from the B-mode image (a) to the skeleton image (d) of the IDC grade III lesion (1). Qualitative comparison of Figure 5 (b) to (d) demonstrate that the main features of top-hat filtered image, such as vessel trunk structure and diameter (b), are preserved in the Hessian-based filtered and binary images.

Based on these results, we estimated the quantitative parameters of vessels (e.g. diameter, number of vessel segments, number of branch points, DM, and SOAM for this image. Figure 6(a) shows the vessel diameter map for the lesion shown in Figure 5 [IDC grade III (1)] where the heat map represents the local diameter variations across different parts of the vascular structure. The remaining Figure 6 images provide the morphological finding for the same lesion. For this image, vessel density is 0.147, the number of vessel segments is 61, and the number of branch points is 28. Average vessel diameter and length with standard deviations are 669.2 ± 232.7 μm and 1603.9 ± 1378.1 μm, respectively. Moreover, the DM and SOAM analysis show their average and standard deviation values to be 1.04 ± 0.09 and 19.08 ± 3.16 deg/μm, respectively.

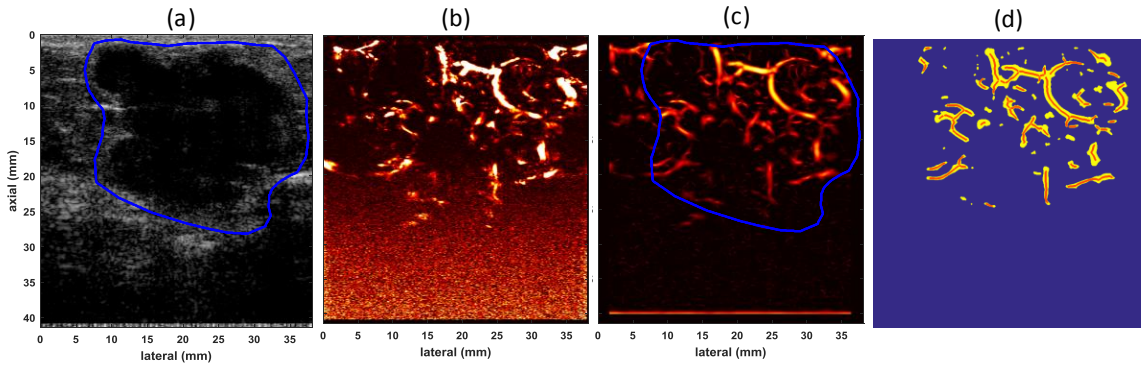

Figure 5(a) Ultrasound B-mode image of an invasive/infiltrating ductal carcinoma (IDC) grade III lesion (1), (b) Microvasculature image after singular value decomposition background noise removal using top-hat filtering and spectral subtraction, (c) Microvasculature image after Hessian-based filtering, and (d) Binary image (yellow) and extracted vessel segments (skeleton of image denoted by red overlay).

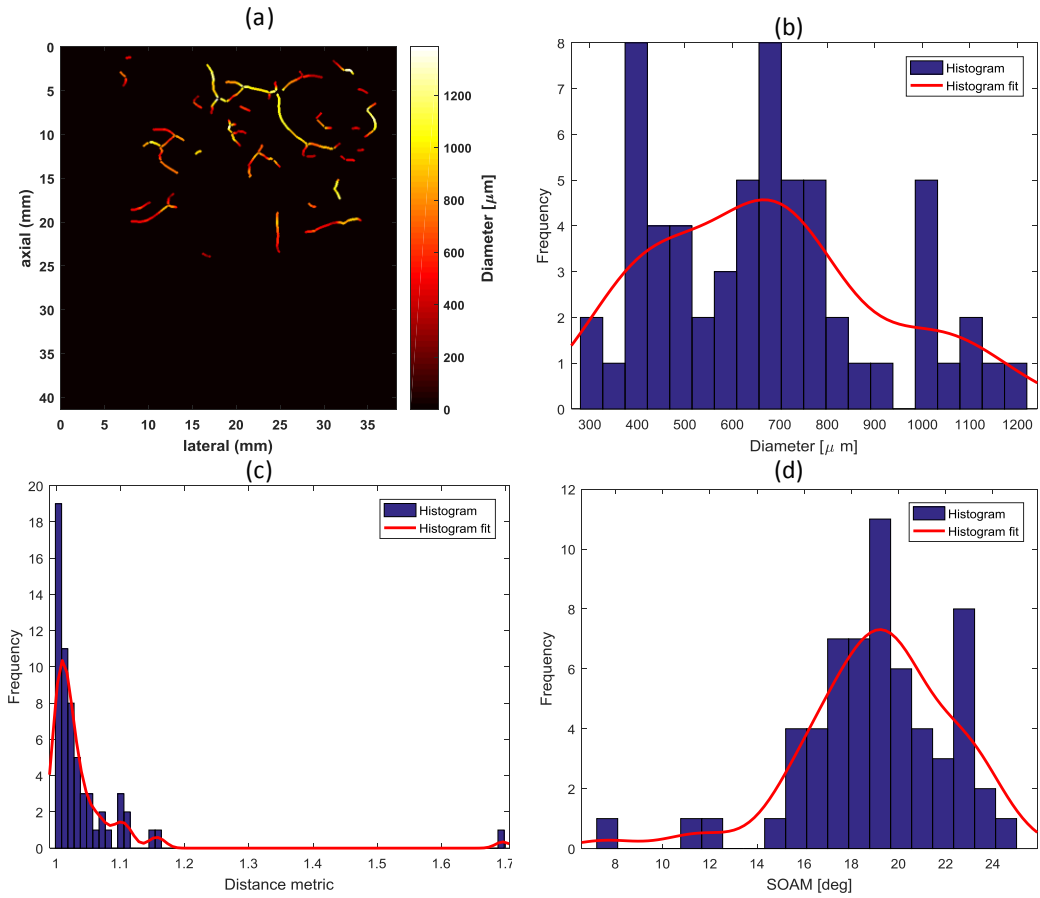

Figure 6 (a) Diameter map of the vessels for Invasive/Infiltrating Ductal Carcinoma (IDC) Grade III lesion (1), (b) Histogram of the diameter, (c) Histogram of the distance metric, and (d) Histogram of sum of angle metric.

The B-mode image of IDC grade III lesion (2) is seen in **Figure 7**(a). A number of representative vessels are seen in Figure 7(b), in which the DM has been applied to highlight the accuracy of measuring tortuosity by DM. As can be observed, the estimated tortuosity well-corroborates with the visual appearance of the vessel segments, such that larger deviations from the straight line indicate increased tortuosity.

To examine the effect of small vessel branches connected to the main vessel trunk on vessel quantification, we studied IDC grade III lesion (2). Figure 8(a) and (b) show the output of the Hessian-based filtering for this lesion with different minimum size scale. As Figure 8(a) shows, there are some branches connected to the main vessel trunk, and these branches cause the main trunk to break down into small vessel segments at branching points. By increasing the minimum size scale of the vessel filter to 462 μm, though, these branching vessels are not visible in Figure 8(b). Hence, the multiple scale size processing capability of the vessel filter enables removing the smaller vessel branches so that large trunks can be more accurately analyzed for tortuosity. The obvious disadvantage of increasing the minimum size scale is losing fine vessel segments with diameters smaller than the minimum allowed. Figure 8(c) and (d) show the corresponding binary images of Figure 8(a) and (b) with extracted vessel segments shown in red color overlaid on the binary segmentation of the vasculature skeleton (yellow). By quantitative comparison of the vessel parameters of Figure 8(a) and (b), the average of the DM increases from 1.047 to 1.071. Average diameter and length of the vessel segments increases, as well, from 588.7 μm and 1154 μm to 658.7 μm and 1776 μm, respectively. Moreover, as we expected, the number of vessel segments decreases from 176 to 83 segments.

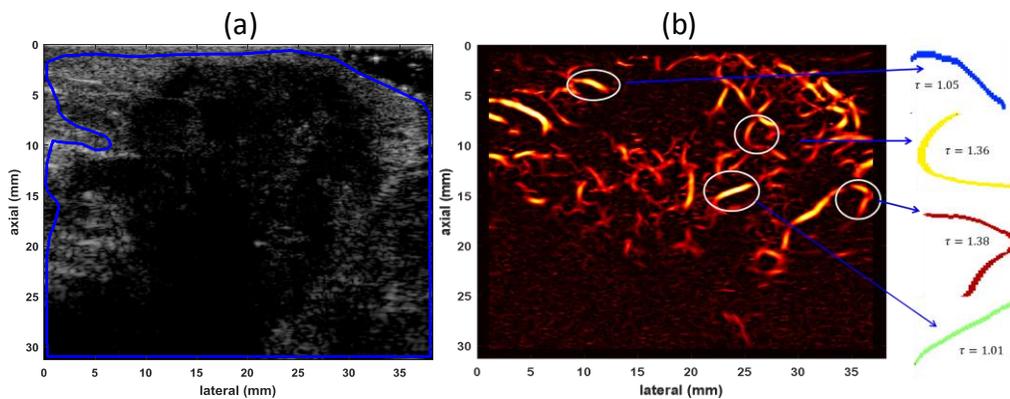

Figure 7. (a) B-mode image of IDC grade III lesion (2), (b) Examples of vessel segments with different levels of distance metric, the white area depicts the vessel where DM of that is calculated

Figure 9(a) shows the mean of the DM [mean(τ)] over different vessel segments of the IDC grade III lesion (2) in Figure 8 as a function of the minimum size scale of the Hessian-based filter (maximum size scale of 500.5 μm). Note that the mean(τ) increases with increasing minimum size scale. The advantage of using a higher value for the minimum size scale is that the contribution of small vessel segments, which may partially appear in the imaging plane, on calculation of the mean(τ) are reduced. Moreover, vessel trunks do not break into smaller vessels, and vessels with larger tortuosity contribute to calculating the mean(τ). Therefore, the mean(τ) does not reduce artificially due to the contribution of partially-appearing vessels in the imaging plane or broken vessel trunks. The mean(τ), though, is not necessarily an increasing function of the minimum size scale, $s$, in different microvasculature images, since vessel trunks may not be tortuous naturally in all microvasculature images. Figure 9(b) shows the mean diameter of the vessel segments as a function of the minimum size scale of the Hessian-based filter. As expected, the mean diameter of vessel segments is an increasing function of the minimum size scale. Figure 9(c) shows the mean length of vessel segment as a function of the minimum size scale. As we expected, due to the removal of the small vessel segments, the mean length of vessel segments is an increasing function of the minimum size scale. Figure 9(d) shows the number of vessel segments as a function of the minimum size scale. It is evident that the number of vessel segments is a decreasing function of the minimum size scale, as fewer branching points are expected to occur when small vessel segments are discarded. Additionally, using higher values for the minimum size scale of vessels for tortuosity analysis provides a more accurate estimation of vessel trunk tortuosity. This effect results from keeping vessel trunks while removing small vessel segments connected to the main vessel trunk, which consequently prevents the breaking of the vessel trunk during the branching procedure.

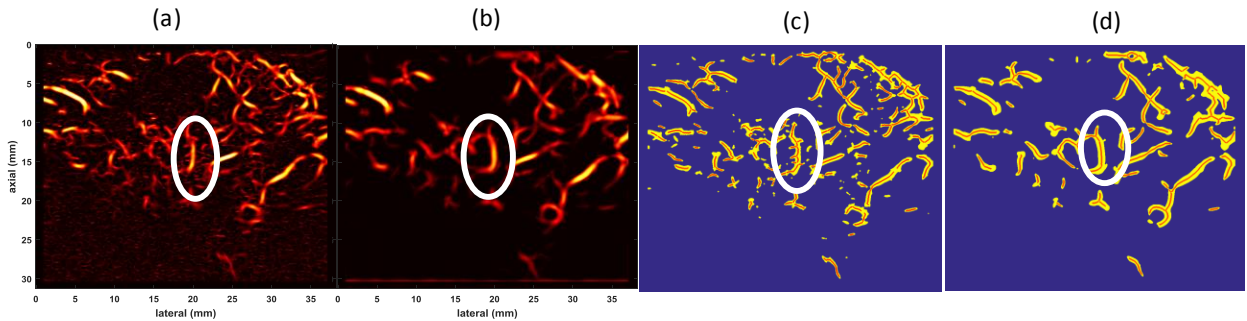

Figure 8 (a) Microvasculature image of invasive/infiltrating ductal carcinoma (IDC) grade III lesion (2) after Hessian based filtering with minimum size scale of 115.5 μm (equivalent to 3 pixels), and (b) after Hessian based filtering with minimum size scale of 423.5 μm (equivalent to 12 pixels). Both (a) and (b) have a maximum size scale of 500.5 μm (equivalent to 15 pixels). (c) Binary image of Figure 8(a) (yellow) with extracted vessel segments (red). (d) Binary image of Figure 8(b) (yellow) with extracted vessel segments (red). The white area depicts branches connected to the main vessel trunk.

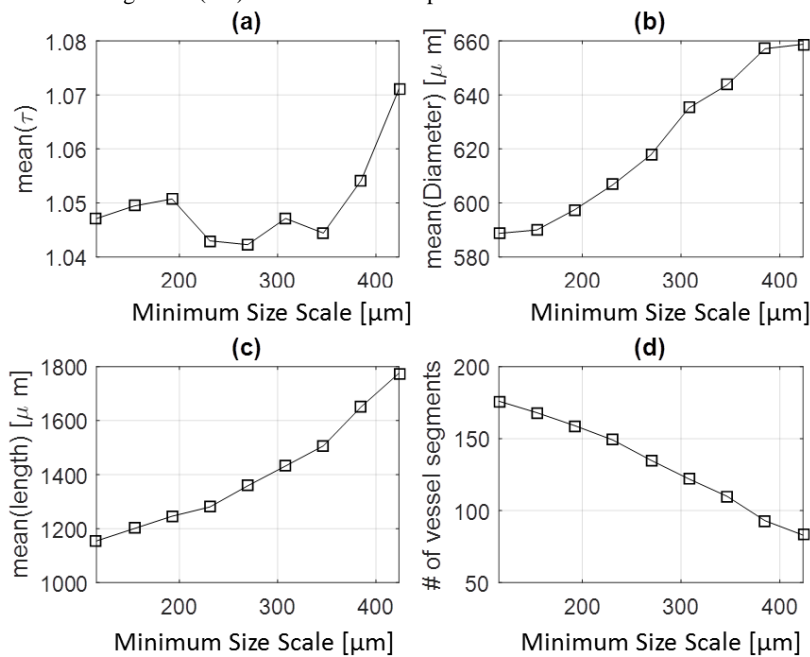

Figure 9. Morphological parameters of the invasive/infiltrating ductal carcinoma (IDC) grade III lesion (2) shown in Figure 8 as a function of the minimum size scale. (a) Mean of the distance metric [mean($\tau$)] over different vessel segments; (b) Mean of the diameter of vessel segments [mean(Diameter)]; (c) Mean of the length of vessel segments; and (d) Number of vessel segments.

To illustrate the effect of erosion-dilatation on removing the small vessel segments connected to a trunk and its impact on the DM, we applied different values for this parameter on the microvasculature image post Hessian-based filtering shown in Figure 8. This process removes small vessel segments connected to the vessel trunk [inside red ellipse in Figure 8(a)]. Figure 10 shows the corresponding binary image of the microvasculature image in Figure 8(a) with a minimum size scale of 115.5 μm and maximum size scale of 346.5 μm for different levels of the erosion-dilation factor. As it can be observed in Figure 10(a-d, inside the white ellipse), by increasing the erosion-dilation factor, small vessel branches disappear while large vessel trunks are preserved in the image; however, at large erosion-dilation factor values, the vessels with small diameters are progressively removed from the image. Additionally, a very large erosion-dilation factor can incur significant distortion in the binary image in comparison with the SVD image.

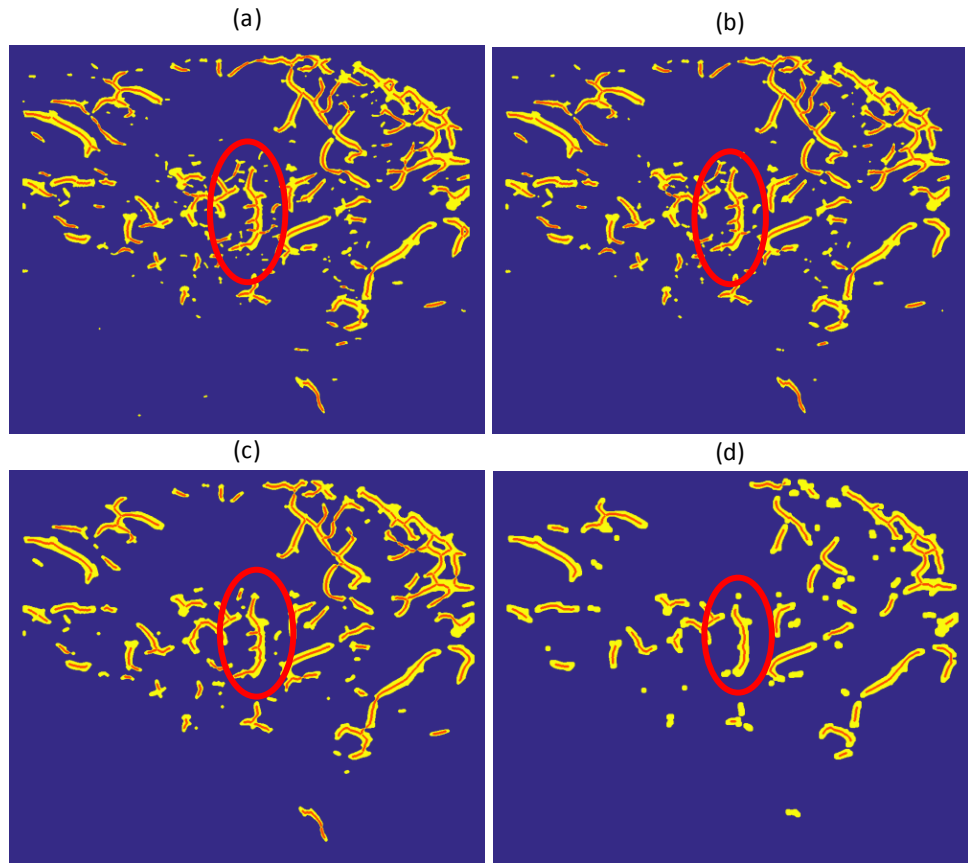

Figure 10. Binary images of the microvascluture image Invasive/Infiltrating Ductal Carcinoma (IDC) Grade III lesion (2) for different levels of the erosion-dilation factor: (a) Without the erosion-dilation factor, (b) With a 77 μm ersion-dialation factor, (c) With a 154 μm erosion-dilation factor, and (d) With a 231 μm erosion-dilation factor.

Figure 11 depicts the mean of the DM [mean(τ)] over different vessel segments of IDC grade III lesion (1) seen in Figure 8(a) as a function of the erosion-dilation (ED) factor in μm. The mean(τ) is shown to be an increasing function of the ED factor. Figure 11(b) demonstrates the mean diameter of vessel segments [mean(diameter)] as a function of the ED factor. As expected, the mean diameter of vessel segments is an increasing function of the ED. Figure 11(c) shows the mean length of vessel segments [mean(length)] as a function of the ED factor . As expected, by removing small vessel segments, the mean length of vessel segments is an increasing function of the ED factor. Figure 11(d) shows the number of vessel segments as a function of the ED. It is evident that the number of vessel segments is a decreasing function of the ED, mainly due to removal of the small vessel segments. By comparison, the results seen in Figure 9(a) and Figure 11(a) demonstrate that using erosion-dilation for removing small vessel segments provides a smoother increment in the DM than changing the minimum size scale of the Hessian filter; however, large

ED values (≥ 6 pixels, equivalent to 231 µm) should be avoided to prevent adding unwanted distortion to the image.

As discussed previously, spectral subtraction may result in a small number of holes in parts of the vessel images where the Doppler shift is perfectly symmetric (i.e., flow perpendicular to the ultrasound beam). These areas need to be filled before further analysis. To assess the performance of our solution for removing these holes with spectral subtraction, we consider two cases. Figure 12(a) and (b) show two vessel segments before removing small holes, and Figure 12(b) and (d) show the corresponding vessel segments after filling small holes, respectively, by dilation followed by erosion. Using this technique can effectively remove small holes in the main vessel trunks with minimal impact on other features of the vessel.

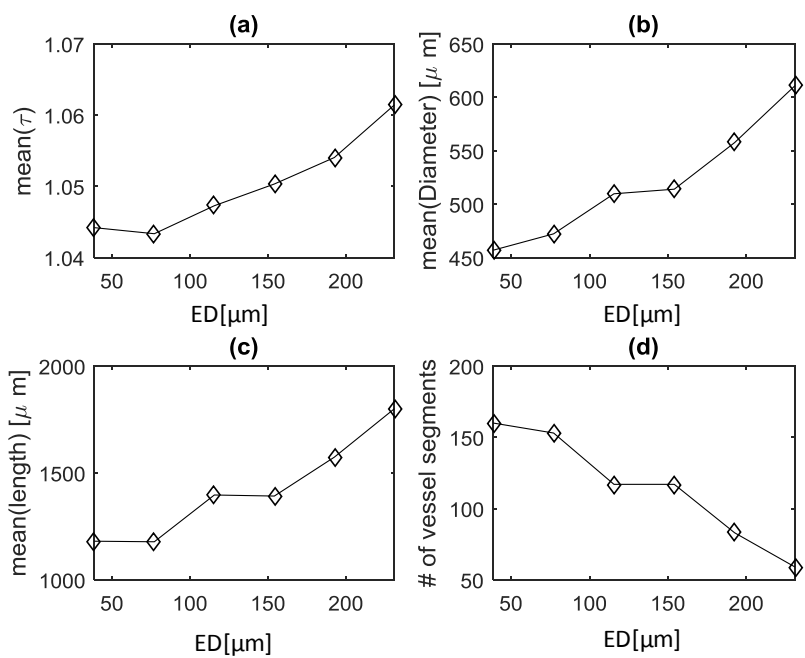

Figure 11. Mean of the distance metric [mean(τ)] over different vessel segments of invasive/infiltrating ductal carcinoma (IDC) grade III lesion(2) in Figure 8(a) in terms of erosion-dilation (ED) factor, (b) Mean of diameter of vessel segments [mean(Diameter)] in terms of ED factor, (c) Mean of length of vessel segments in terms of ED factor, and (d) Number of vessel segments in terms of ED factor.

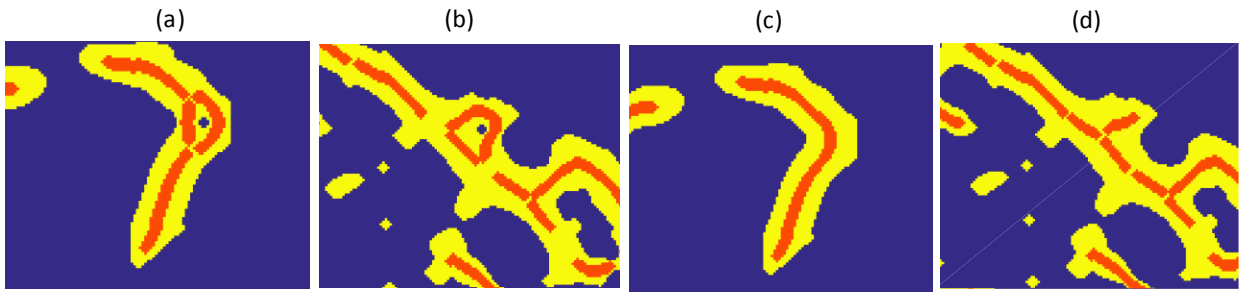

Figure 12(a,b) Vessel segment before removing small hole (c,d) Vessel segment corresponding to (a,b) after removing small holes.

To demonstrate potential diagnostic applications of micro vessel quantification, it is helpful to study examples of vessel quantification applied to in vivo data from benign and malignant breast masses. The second case is the invasive/infiltrating ductal carcinoma (IDC) grade III lesion (2) shown in Figure 8. The estimated parameters of the vessel segments for the microvasculature image shown in Figure 8 (minimum size scale 115.5 μm, maximum of 346.5 μm, consistent with size scale dimensions in other lesions), are shown in Figure 13. From the normalized histogram of distance metric for that lesion; it can be seen there are many vessel segments with distance metric of 1. For this lesion, the vessel density is 0.1154, number of vessel segments is 153, and number of branch points is 101. The respective average and standard deviations of vessel diameter and vessel length are 472.2 ± 149.8 μm and 1178.0 ± 811.6 μm. Moreover, DM and SOAM analysis shows the average and standard deviation values of DM and SOAM are 1.043 ± 0.0629 and 22.8 ± 3.05deg/μm, respectively.

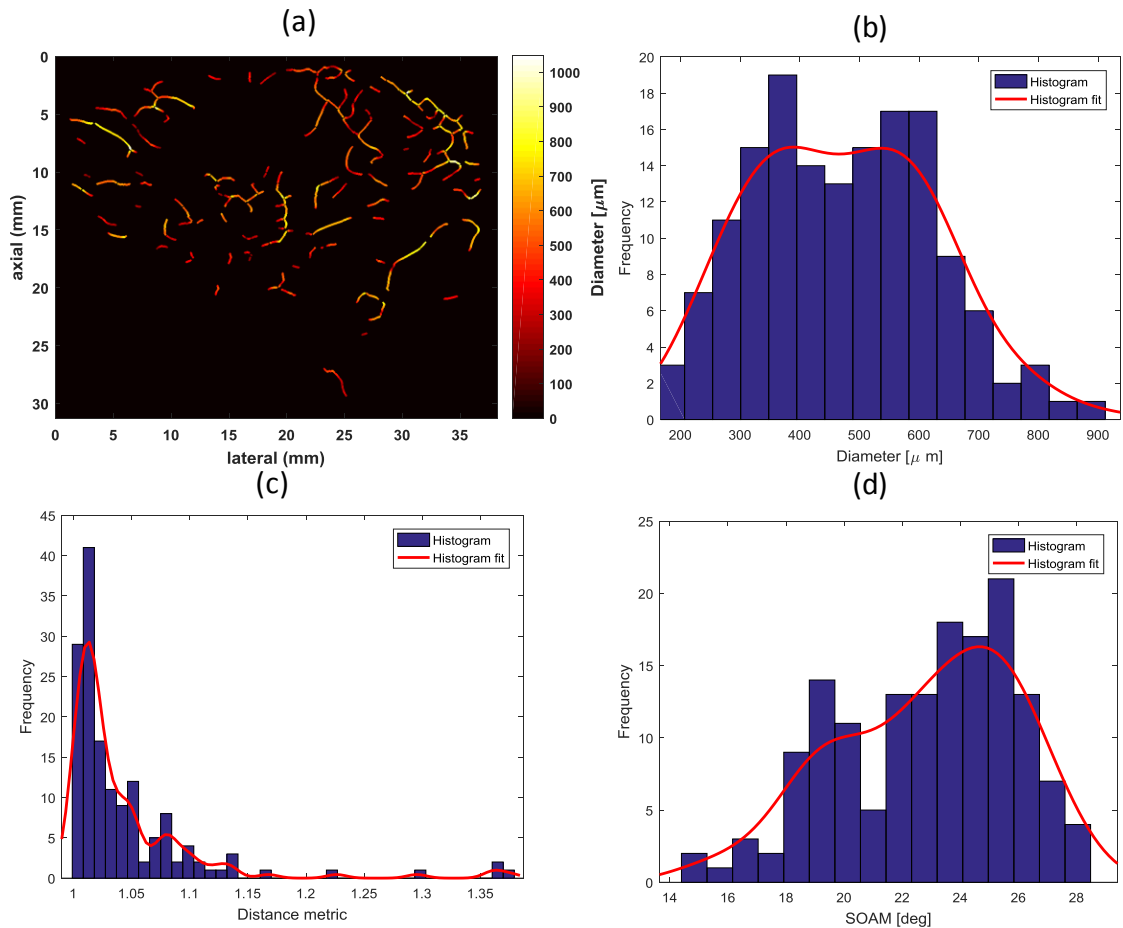

Figure 13 (a) Diameter map of vessels for invasive/infiltrating ductal carcinoma (IDC) grade III lesion (2), (b) Histogram of diameter, (c) Histogram of distance metric, and (d) Histogram of sum of angle metric.

Figure 14 presents the quantification of the vasculature for the next study case: a biopsy-proven fibroadenoma (3). Results of the morphological analysis summarized in Figure 15. For this lesion, vessel density is 0.21, number of vessel segments is 30, and the number of branch points is 15. Averages and standard deviations of vessel diameter and vessel length are 435.2 ± 115.9 μm and 1561.9 ± 1051.3 μm, respectively. Moreover, DM and SOAM analyses show their average values to be 1.083 ± 0.18 and 18.37 ± 2.29 deg/μm, respectively.

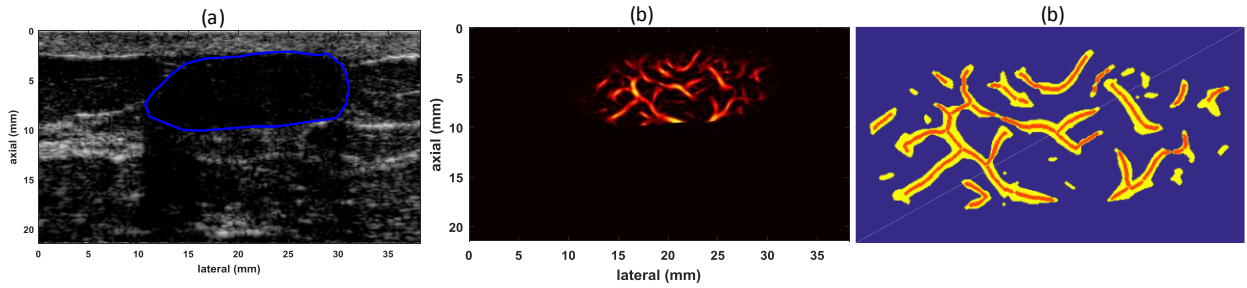

Figure 14 (a) Ultrasound B-mode image of of fibroadenoma lesion (3), (b) Microvascluture imag of fibroadenoma lesion after hessian-based filtering , and (c) Binary image of microvascluture image (yellow) with 77 μm of erosion-dilation factor and extracted vessel segments(red).

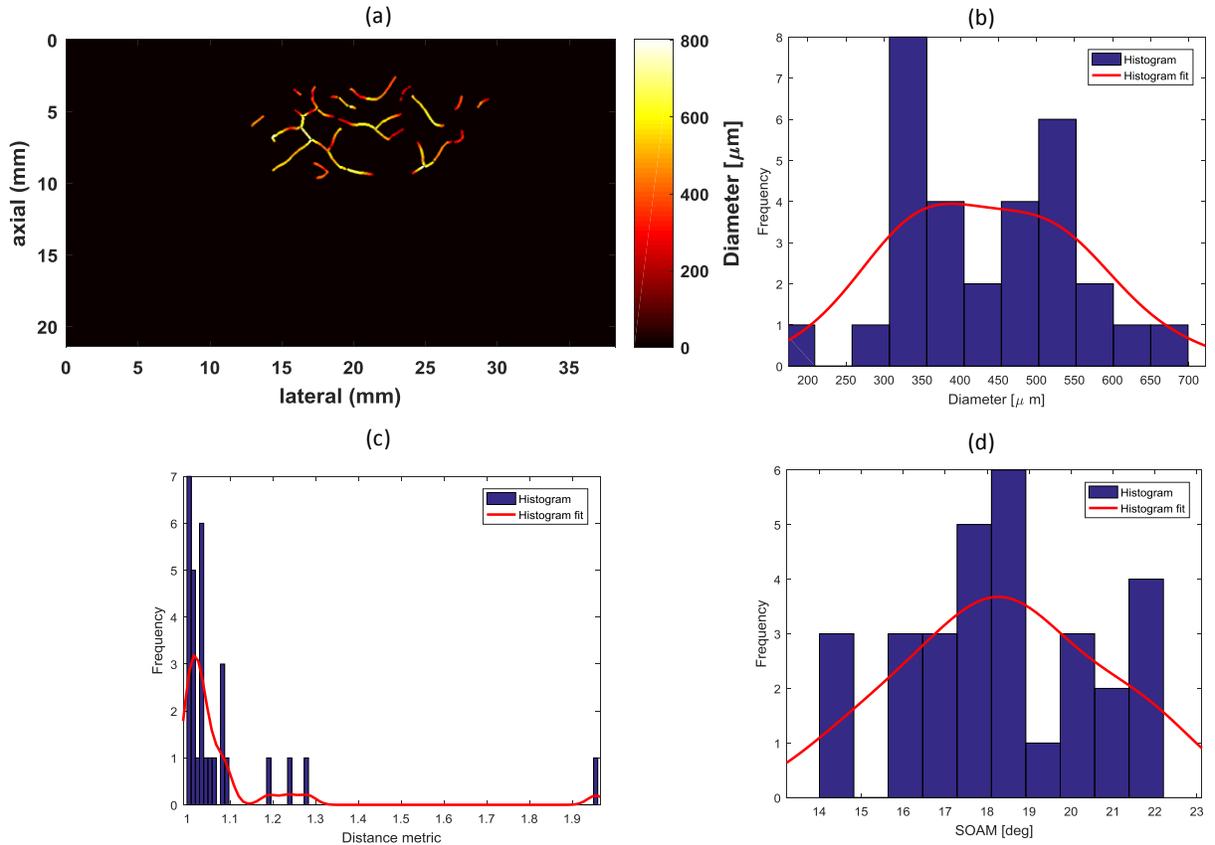

Figure 15 (a) Diameter map of vessels for benign fibroadenoma lesion (3), (b) Histogram of diameter, (c) Distance metric map of vessels, (d) Histogram of distance metric, (e) Sum of angle metric(SOAM) map, and (f) Histogram of sum of angle metric.

The last case is a highly vascularized fibroadenoma breast lesion (4) shown in Figure 16. For this image vessel density is 0.105; the number of vessel segments is 24; the number of branch points is 15. Results of the morphological analysis summarized in Figure 17. Average vessel diameter and length are 385.26 ± 88.8 μm and 1221.5 ± 1117.6 μm, respectively. Moreover, DM and SOAM analyses show the average and standard deviation values to be 1.03 ± 0.07 and 16.95 ± 4.65 deg/μm, respectively.

Here, we study the effect of different threshold values on minimum acceptable vessel length. This threshold can remove isolated vessel segments but cannot keep vessel trunks. By removing isolated vessel segments, we found it possible to increase the average tortuosity in the image. Figure 18 shows the number of vessel segments, mean of vessel segment length, mean of vessel segment diameter, and mean of DM in terms of threshold, defining the minimum vessel length for our four examined lesions. As expected, by increasing the threshold for minimum vessel length, the number of vessel segments is reduced and the mean of vessel segment length is increased, since by increasing the minimum threshold of vessel length, more small vessel segments are removed and longer vessels remain for analysis. Vessel segment diameter remains almost constant in terms of the threshold of minimum vessel length. Also, DM exhibits an increasing behavior in terms of the threshold of minimum vessel length, since small vessel segments normally are straight; therefore mean ($\tau$) increases by removing small vessel segments.

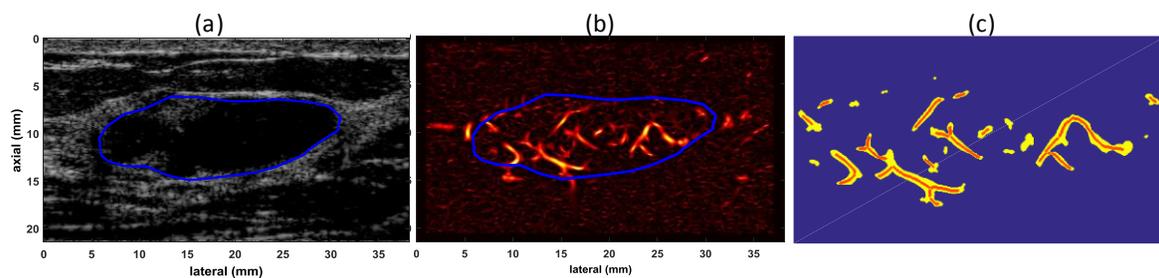

Figure 16 (a) Ultrasound B-mode image of benign fibro adenoma lesion (4), (b) Microvascluture image of fibroadenoma lesion, and (c) Binary image of microvascluture image (yellow) and extracted vessel segments (red).

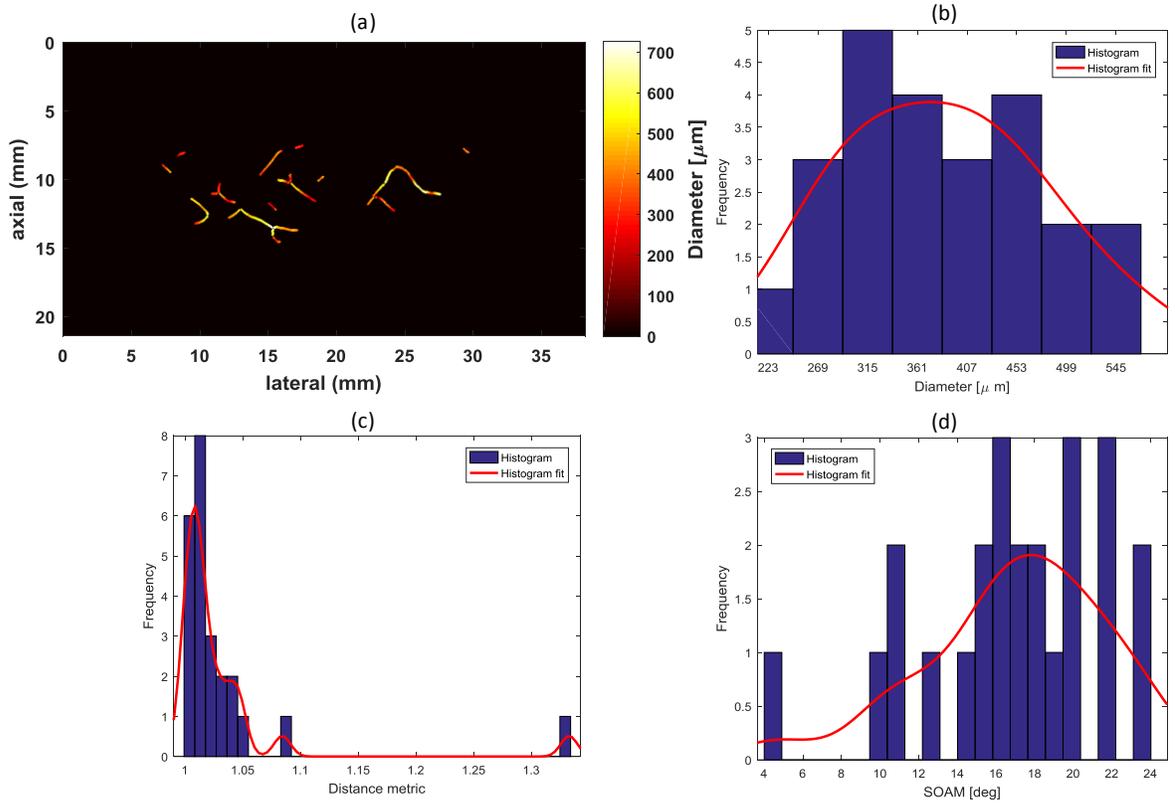

Figure 17 (a) Diameter map of vessels for benign fibroadenoma lesion (4) (b) histogram of diameter (c) histogram of distance metric (d) Sum of angle metric(SOAM) map (f) histogram of sum of angle metric.

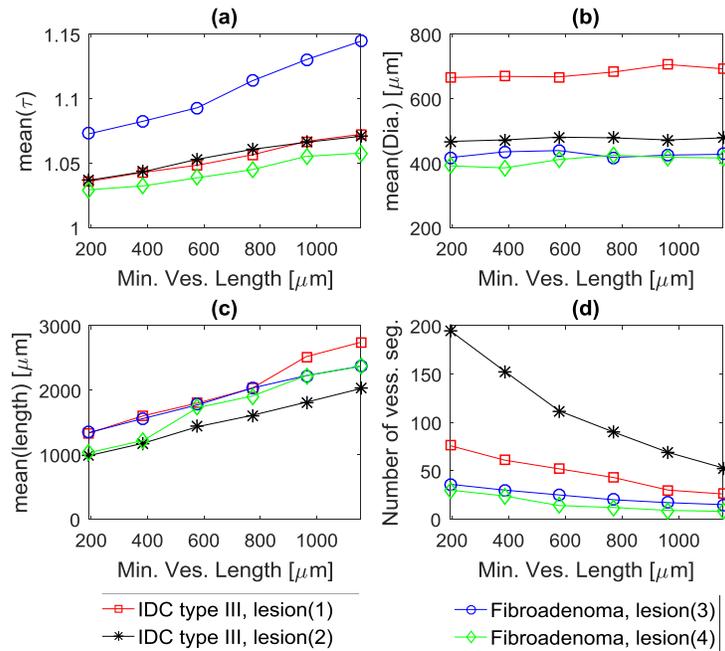

Figure 18(a) Mean of distance metric, (b) Mean of vessel segment diameter, (c) Mean of vessel segment length, and (d) Number of vessel segments in terms of threshold of minimum length for four different patients, two with IDC grade III, and two with Benign fibroadenoma.

Quantitative assessment of these four lesions is summarized in Table 2. Of all the lesions, fibroadenoma lesion (3) has the largest vessel density (0.21), whereas the density found in fibroadenoma lesion (4) is similar to those of the malignant lesions. The number of vessel segments and branch points in the benign lesions are smaller than those of the malignant lesions. Mean and standard deviation vessel diameter values are smaller in the fibroadenoma lesions in comparison with those values from the IDCs. Moreover, these two malignant cases have higher SOAM values than the benign.

Table 2. Estimated quantification parameters for four different lesions..

|  | IDC grade III lesion (1) | IDC grade III lesion(2) | Fibroadenoma lesion (3) | Fibroadenoma lesion (4) |
|---|---|---|---|---|
| Vessel density | 0.15 | 0.12 | 0.21 | 0.10 |
| Mean($\tau$) | 1.04±0.09 | 1.04±0.06 | 1.08±0.18 | 1.03±0.070 |
| Median($\tau$) | 1.02 | 1.02 | 1.03 | 1.01 |
| Max($\tau$) | 1.70 | 1.38 | 1.96 | 1.33 |
| Mean(SOAM) [deg/μm] | 19.08±3.16 | 22.84±3.08 | 18.37±2.29 | 16.95±4.65 |
| Median(SOAM)[deg/μm] | 19.40 | 23.46 | 18.46 | 17.50 |
| Max(SOAM) [deg] | 24.88 | 28.40 | 22.12 | 23.86 |
| Mean(Diameter) [μm] | 669.2±232.7 | 472.2±149.9 | 435.2±115.9 | 385.3±88.8 |
| Median(Diameter) [μm] | 658.0 | 481.7 | 424.8 | 381.1 |
| Max(Diameter) [μm] | 1216.9 | 912.0 | 697.0 | 567.9 |
| Mean(length) [μm] | 1603.9±1378.1 | 1178.0±811.6 | 1561.9±1051.3 | 1221.5±1117.6 |
| Median(length) [μm] | 1076.5 | 968.4 | 1273.8 | 771.5 |
| Max(length) [μm] | 6852.7 | 5033.5 | 4181.3 | 5453.3 |
| Number of vessel segments | 61 | 153 | 30 | 24 |
| Number of branch points | 28 | 101 | 15 | 15 |

## IV. CONCLUSIONS

A set of methods for quantification of the tissue microvasculature obtained by non-contrast ultrasonic microvasculature imaging was presented. The micro-vasculature map comprises vessel segments resulting from blood activity. We introduced procedures to acquire morphometric parameters with additional morphological constrains to reduce erroneous data. Vascular structures were accepted as vessel segments when multiple constraints on amplitude of the vessel segment, diameter of the vessel segment, and length of the vessel segment were satisfied. We addressed challenges in acquiring segmentation-ready microvasculature images and showed that a combination of background removal and vessel enhancement filtering allows vessel segmentation and skeletonization, in turn allowing morphological analysis. The quantitative parameters may include tortuosity measures, (DM and SOAM), diameter of vessel segments, length of vessel segments, number of vessel segments, number of branching points, and vessel density. Given the 2-D nature of B-mode ultrasound imaging, accurate interpretation of some vascular features can be difficult. 2-D cross sectional imaging may provide erroneous branching and vessel crossings that may lead to incorrect interpretation of the vessel segments. While quantitative evaluation of parameters, such as vessel density and diameter, are not significantly affected by this phenomenon, measures of the tortuosity, number of branching points, and number of vessel segments may become inaccurate. In this paper, we introduced a number of strategies to enable extraction of several morphological features by adding additional constrains. The most important contribution was to devise methods to preserve large vessel trunks that may be broken into small pieces due to intersection with out-of-plane vessel segments, namely by removing small size-scales from the vessel filtering and small vessel segments connected to large trunks via morphological operations. Another limitation in ultrasonic microvasculature images is related to small vessel segments which may result from cross sectional imaging of vessels. These may appear as small vessel segments with incorrect information regarding the vascular tree segments. We addressed this issue by enforcing vessel segment length and diameter constraints. The methods presented in this paper provide a set of tools for quantitative assessment of microvasculature morphological features. These features may be associated with certain diseases or different health conditions. In cancer, for example, malignant tumors have been shown to give rise to tortuous vessels. The initial results in this paper suggest the quantitative morphological parameters may allow differentiation of certain lesions, such as benign and malignant breast lesions. Therefore, the methods presented in this paper for quantitative assessment of microvasculature morphological features obtained from non-contrast ultrasound images may result in potential biomarkers for the diagnosis of some diseases.

## ACKNOWLEDGMENT

This work was supported by the NIH Grants R01EB017213, R01CA148994, R01CA168575, R01CA195527, and Ro1CA174723. The authors wish to thanks Desiree J. Lanzino, PT, PhD, for her help in editing the paper.

## V. REFERENCES

[1] D. Ribatti, B. Nico, S. Ruggieri, R. Tamma, G. Simone, and A. Mangia, "Angiogenesis and Antiangiogenesis in Triple-Negative Breast cancer," *Translational Oncology,* vol. 9, no. 5, pp. 453-457.
[2] B. P. Schneider and K. D. Miller, "Angiogenesis of breast cancer," (in eng), *J Clin Oncol,* vol. 23, no. 8, pp. 1782-90, Mar 10 2005.
[3] B. R. Zetter, "Angiogenesis and tumor metastasis," (in eng), *Annu Rev Med,* vol. 49, pp. 407-24, 1998.
[4] E. Bullitt *et al.*, "Tumor therapeutic response and vessel tortuosity: preliminary report in metastatic breast cancer," (in eng), *Med Image Comput Comput Assist Interv,* vol. 9, no. Pt 2, pp. 561-8, 2006.
[5] A. H. Parikh, J. K. Smith, M. G. Ewend, and E. Bullitt, "Correlation of MR perfusion imaging and vessel tortuosity parameters in assessment of intracranial neoplasms," (in eng), *Technol Cancer Res Treat,* vol. 3, no. 6, pp. 585-90, Dec 2004.
[6] E. Bullitt, G. Gerig, S. M. Pizer, L. Weili, and S. R. Aylward, "Measuring tortuosity of the intracerebral vasculature from MRA images," *IEEE Transactions on Medical Imaging,* vol. 22, no. 9, pp. 1163-1171, 2003.
[7] W. E. Hart, M. Goldbaum, B. Côté, P. Kube, and M. R. Nelson, "Measurement and classification of retinal vascular tortuosity," *International Journal of Medical Informatics,* vol. 53, no. 2, pp. 239-252, 1999/02/01/ 1999.
[8] J. Ehling *et al.*, "Micro-CT Imaging of Tumor Angiogenesis: Quantitative Measures Describing Micromorphology and Vascularization," *The American journal of pathology,* vol. 184, no. 2, pp. 431-441, 11/18 2014.
[9] R. C. Gessner, S. R. Aylward, and P. A. Dayton, "Mapping microvasculature with acoustic angiography yields quantifiable differences between healthy and tumor-bearing tissue volumes in a rodent model," (in eng), *Radiology,* vol. 264, no. 3, pp. 733-40, Sep 2012.
[10] S. E. Shelton *et al.*, "Quantification of Microvascular Tortuosity during Tumor Evolution Using Acoustic Angiography," *Ultrasound in Medicine and Biology,* vol. 41, no. 7, pp. 1896-1904, 2015.
[11] F. Lin, S. E. Shelton, D. Espindola, J. D. Rojas, G. Pinton, and P. A. Dayton, "3-D Ultrasound Localization Microscopy for Identifying Microvascular Morphology Features of Tumor Angiogenesis at a Resolution Beyond the Diffraction Limit of Conventional Ultrasound," (in eng), *Theranostics,* vol. 7, no. 1, pp. 196-204, 2017.
[12] S. K. Kasoji, J. N. Rivera, R. C. Gessner, S. X. Chang, and P. A. Dayton, "Early Assessment of Tumor Response to Radiation Therapy using High-Resolution Quantitative Microvascular Ultrasound Imaging," (in eng), *Theranostics,* vol. 8, no. 1, pp. 156-168, 2018.
[13] B. Theek, T. Opacic, T. Lammers, and F. Kiessling, "Semi-Automated Segmentation of the Tumor Vasculature in Contrast-Enhanced Ultrasound Data," *Ultrasound in Medicine & Biology,* vol. 44, no. 8, pp. 1910-1917, 2018/08/01/ 2018.
[14] S. R. Wilson and P. N. Burns, "Microbubble-enhanced US in Body Imaging: What Role?," *Radiology,* vol. 257, no. 1, pp. 24-39, 2010.
[15] J. D. Thomas, "Myocardial Contrast Echocardiography Perfusion Imaging," *Still Waiting After All These Years,* vol. 62, no. 15, pp. 1362-1364, 2013.
[16] R. Gessner and P. A. Dayton, "Advances in Molecular Imaging with Ultrasound," *Molecular imaging,* vol. 9, no. 3, pp. 117-127, 2010.
[17] J. R. Lindner, "Molecular imaging of cardiovascular disease with contrast-enhanced ultrasonography," *Nature Reviews Cardiology,* Review Article vol. 6, p. 475, 06/09/online 2009.
[18] F. Molinari, A. Mantovani, M. Deandrea, P. Limone, R. Garberoglio, and J. S. Suri, "Characterization of Single Thyroid Nodules by Contrast-Enhanced 3-D Ultrasound," *Ultrasound in Medicine & Biology,* vol. 36, no. 10, pp. 1616-1625, 2010/10/01/ 2010.
[19] S. F. Huang, R. F. Chang, W. K. Moon, Y. H. Lee, D. R. Chen, and J. S. Suri, "Analysis of Tumor Vascularity Using Three-Dimensional Power Doppler Ultrasound Images," *IEEE Transactions on Medical Imaging,* vol. 27, no. 3, pp. 320-330, 2008.

[20] J. R. Eisenbrey, N. Joshi, J. K. Dave, and F. Forsberg, "Assessing algorithms for defining vascular architecture in subharmonic images of breast lesions," (in eng), *Phys Med Biol,* vol. 56, no. 4, pp. 919-30, Feb 21 2011.

[21] C. Demené *et al.*, "Spatiotemporal Clutter Filtering of Ultrafast Ultrasound Data Highly Increases Doppler and fUltrasound Sensitivity," *IEEE Transactions on Medical Imaging,* vol. 34, no. 11, pp. 2271-2285, 2015.

[22] M. Bayat, M. Fatemi, and A. Alizad, "Background Removal and Vessel Filtering of Non-Contrast Ultrasound Images of Microvasculature," *IEEE Transactions on Biomedical Engineering,* pp. 1-1, 2018.

[23] J. Bercoff *et al.*, "Ultrafast compound doppler imaging: providing full blood flow characterization," *IEEE Transactions on Ultrasonics, Ferroelectrics, and Frequency Control,* vol. 58, no. 1, pp. 134-147, 2011.

[24] E. Cohen, T. Deffieux, E. Tiran, C. Demene, L. Cohen, and M. Tanter, "Ultrasensitive Doppler based neuronavigation system for preclinical brain imaging applications," in *2016 IEEE International Ultrasonics Symposium (IUS)*, 2016, pp. 1-4.

[25] M. M. Fraz *et al.*, "Blood vessel segmentation methodologies in retinal images – A survey," *Computer Methods and Programs in Biomedicine,* vol. 108, no. 1, pp. 407-433, 2012/10/01/ 2012.

[26] N. Passat, C. Ronse, J. Baruthio, J. P. Armspach, C. Maillot, and C. Jahn, "Region-growing segmentation of brain vessels: an atlas-based automatic approach," (in eng), *J Magn Reson Imaging,* vol. 21, no. 6, pp. 715-25, Jun 2005.

[27] T. Boskamp, D. Rinck, F. Link, B. Kümmerlen, G. Stamm, and P. Mildenberger, "New Vessel Analysis Tool for Morphometric Quantification and Visualization of Vessels in CT and MR Imaging Data Sets," *RadioGraphics,* vol. 24, no. 1, pp. 287-297, 2004.

[28] S. Moccia, E. De Momi, S. El Hadji, and L. S. Mattos, "Blood vessel segmentation algorithms — Review of methods, datasets and evaluation metrics," *Computer Methods and Programs in Biomedicine,* vol. 158, pp. 71-91, 2018/05/01/ 2018.

[29] E. Cohen, L. D. Cohen, T. Deffieux, and M. Tanter, "An Isotropic Minimal Path Based Framework for Segmentation and Quantification of Vascular Networks," Cham, 2018, pp. 499-513: Springer International Publishing.

[30] E. J. Cand, X. Li, Y. Ma, and J. Wright, "Robust principal component analysis?," *J. ACM,* vol. 58, no. 3, pp. 1-37, 2011.

[31] O. Ricardo, C. Emmanuel, and S. D. K., "Low-rank plus sparse matrix decomposition for accelerated dynamic MRI with separation of background and dynamic components," *Magnetic Resonance in Medicine,* vol. 73, no. 3, pp. 1125-1136, 2015.

[32] G. Hao, Y. Hengyong, O. Stanley, and W. Ge, "Multi-energy CT based on a prior rank, intensity and sparsity model (PRISM)," *Inverse Problems,* vol. 27, no. 11, p. 115012, 2011.

[33] S. G. Lingala, Y. Hu, E. DiBella, and M. Jacob, "Accelerated Dynamic MRI Exploiting Sparsity and Low-Rank Structure: k-t SLR," *IEEE Transactions on Medical Imaging,* vol. 30, no. 5, pp. 1042-1054, 2011.

[34] G. Strang, *Introduction to Linear Algebra*, 5th ed. Wellesley MA: Wellesley-Cambridge Press, My 2016.

[35] E. R. Dougherty, *An introduction to morphological image processing*. SPIE Optical Engineering Press, Feb. 1992.

[36] A. F. Frangi, W. J. Niessen, K. L. Vincken, and M. A. Viergever, "Multiscale vessel enhancement filtering," Berlin, Heidelberg, 1998, pp. 130-137: Springer Berlin Heidelberg.

[37] J. Serra, "Introduction to mathematical morphology," *Comput. Vision Graph. Image Process.,* vol. 35, no. 3, pp. 283-305, 1986.

[38] L. Lam, S. W. Lee, and C. Y. Suen, "Thinning methodologies-a comprehensive survey," *IEEE Transactions on Pattern Analysis and Machine Intelligence,* vol. 14, no. 9, pp. 869-885, 1992.

[39] R. C. Gonzalez, R. E. Woods, and S. L. Eddins, *Digital Image Processing Using MATLAB*. Prentice-Hall, Inc., 2003.